\documentclass[aps,pre,superscriptaddress,amsmath,amssymb,superscriptaddress,twocolumn,floatfix,showpacs,longbibliography]{revtex4-1}

\usepackage{graphicx}
\usepackage{dcolumn}
\usepackage{bm}
\begin{document}
\title{Collective dynamics of pedestrians interacting with attractions}

\author{Jaeyoung Kwak}
\email{jaeyoung.kwak@aalto.fi}
\affiliation{Department of Civil and Environmental Engineering, Aalto University School of Engineering, P.O. Box 12100, FI-00076, Finland}
\author{Hang-Hyun Jo}
\affiliation{BECS, Aalto University School of Science, P.O. Box 12200, FI-00076, Finland}
\author{Tapio Luttinen}
\affiliation{Department of Civil and Environmental Engineering, Aalto University School of Engineering, P.O. Box 12100, FI-00076, Finland}
\author{Iisakki Kosonen}
\affiliation{Department of Civil and Environmental Engineering, Aalto University School of Engineering, P.O. Box 12100, FI-00076, Finland}

\date{December 10, 2013}

\begin{abstract}
In order to investigate collective effects of interactions between pedestrians and attractions, this study extends the social force model. Such interactions lead pedestrians to form stable clusters around attractions, or even to rush into attractions if the interaction becomes stronger. It is also found that for high pedestrian density and intermediate interaction strength, some pedestrians rush into attractions while others move to neighboring attractions. These collective patterns of pedestrian movements or phases and transitions between them are systematically presented in a phase diagram. The results suggest that safe and efficient use of pedestrian areas can be achieved by moderating the pedestrian density and the strength of attractive interaction, for example, in order to avoid situations involving extreme desire for limited resources.
\end{abstract}

\pacs{89.40.-a, 89.65.-s, 05.65.+b}

\maketitle
\section{Introduction}
Collective patterns of pedestrian movements have received increasing attention from various fields including physics [1], transportation engineering [2], and marketing [3]. In an attempt to understand mechanisms underlying collective patterns, different microscopic pedestrian behavior models have been proposed, such as cellular automata (CA) [4,5], social force models [6–8], and centrifugal force models [9,10]. Among these models, the social force model has produced promising results in that the model embodies behavioral elements and physical forces. The original model and its variants have successfully demonstrated various interesting phenomena such as lane formation [6], bottleneck oscillation [6], and turbulent movement [11]. Such self-organized patterns could be understood to some extent by considering mainly repulsive interactions among pedestrians.

Collective patterns of pedestrian movements have received increasing attention from various fields including physics~\cite{Helbing_RMP2001}, transportation engineering~\cite{Daamen_TRR2003}, and marketing~\cite{Hui_MarSci2009}. In an attempt to understand mechanisms underlying collective patterns, different microscopic pedestrian behavior models have been proposed, such as cellular automata (CA)~\cite{Blue_2001, Burstedd_2001}, social force models~\cite{Helbing_PRE1995, Helbing_TrSci2005, Koster_PRE2013}, and centrifugal force models~\cite{Yu_PRE2005, Chraibi_PRE2010}. Among these models, the social force model has produced promising results in that the model embodies behavioral elements and physical forces. The original model and its variants have successfully demonstrated various interesting phenomena such as lane formation~\cite{Helbing_PRE1995}, bottleneck oscillation~\cite{Helbing_PRE1995}, and turbulent movement~\cite{Yu_PRE2009}. Such self-organized patterns could be understood to some extent by considering mainly repulsive interactions among pedestrians.

In addition to the repulsive interactions, the effect of attractive interactions also has been investigated, which was first introduced by Helbing and Moln\'{a}r~\cite{Helbing_PRE1995}. For example, Xu and Duh~\cite{Xu_IEEE2010} simulated bonding effects of pedestrian groups and evaluated their impacts on pedestrian flows. In another study, Moussa\"{i}d \textit{et al.}~\cite{Moussaid_PLOS2010} analyzed attractive interactions among pedestrian group members and their spatial patterns. 

Although several studies focused on attractive interactions among pedestrians, little attention has been paid to the attractive interactions between pedestrians and attractions. While walking to the destinations, pedestrians can be influenced by unexpected attractions such as window displays and street performances. If the attractive force between pedestrians and attractions is strong enough, pedestrians may stop walking to destinations and join attractions, namely impulse stops~\cite{Borgers_1986, Timmermans_1992}. For instance, museum managers have concerned with stop patterns of visitors because impulse stops in museums initiate contacts between visitors and museum displays~\cite{Serrell_1997}. Thus, architects have proposed various museum layouts in order to motivate visitors to have contact with more museum displays~\cite{Rohloff_2009}. In the marketing area, since impulse stops in stores may result in unplanned purchases, marketing strategies have focused on exposing consumers to more merchandise and encouraging them to stop to buy~\cite{Hui_JMR_2013}. Understanding such impulse stops can support effective design and management of pedestrian facilities.  

In order to examine the collective effects of attractive interactions between pedestrians and attractions, this study extends the social force model by incorporating the attractive interactions. The extended social force model reproduces various collective patterns of pedestrian movements or phases, depending on the strength of attractive interaction and the pedestrian density. These phases are systematically presented in a phase diagram. 

This paper is organized as follows. The extended social force model is described in Sec.~\ref{sec:model}, and its numerical simulation results are presented in terms of a phase diagram in Sec.~\ref{sec:Results}. Finally, Sec.~\ref{sec:summary} summarizes the results with concluding remarks. 

\section{Model}
\label{sec:model}
The social force model~\cite{Helbing_PRE1995} describes the pedestrian movements in terms of the superposition of driving, repulsive, and attractive force terms, analogously to self-propelled particle models~\cite{Vicsek_PRL1995, Mogilner_MathBio2003, DOrsogna_PRL2006, Silverberg_PRL2013}. Each pedestrian $i$ is modeled as a circle with radius $r_i$ in a two-dimensional space. The position and velocity of each pedestrian $i$ at time $t$, denoted by $\vec x_i(t)$ and $\vec v_i(t)$, evolve according to the following equations:
\begin{equation}
\frac{d\vec{x}_i(t)}{dt} =\vec v_i(t)
\end{equation}
and
\begin{eqnarray}\label{eq:EoM}
\frac{d\vec{v}_i(t)}{dt} = \vec{f}_{i,d}+\sum_{j\neq i}^{ }{\vec{f}_{ij}}+\sum_{B}^{ }{\vec{f}_{iB}}+\sum_{A}^{ }{\vec{f}_{iA}}.
\end{eqnarray}
Here the driving force $\vec{f}_{i,d}$ describes the pedestrian $i$ accelerating to reach its destination. The repulsive force between pedestrians $i$ and $j$, $\vec{f}_{ij}$, denotes the tendency of pedestrians to keep a certain distance from each other. The repulsive force from boundaries $\vec{f}_{iB}$ shows the interaction between pedestrians and boundaries, e.g., wall and obstacles. The attractive force $\vec{f}_{iA}$ indicates pedestrian movements toward attractive stimuli, e.g., window displays and museum exhibits. 

The driving force $\vec{f}_{i,d}$ is given as
\begin{eqnarray}\label{eq:driving}
\vec{f}_{i,d} = \frac{v_d\vec{e}_{i}-\vec{v}_{i}(t)}{\tau},
\end{eqnarray}
where ${v}_{d}$ is the desired speed and $\vec{e}_{i}$ is an unit vector for the desired direction, independent of the position of the pedestrian $i$. The relaxation time $\tau$ controls how fast the pedestrian $i$ adapts its velocity to the desired velocity. The repulsive force between pedestrians $i$ and $j$, denoted by $\vec{f}_{ij}$, is the sum of the gradient of repulsive potential with respect to $\vec d_{ij}\equiv \vec x_j-\vec x_i$, and the friction force, $\vec g_{ij}$: 
\begin{eqnarray}\label{eq:repulsive1}
  \vec{f}_{ij} &=& -\nabla_{\vec{d}_{ij}}V(b_{ij})+\vec g_{ij}.
\end{eqnarray}
The repulsive potential is given as
\begin{eqnarray}
  V(b_{ij}) &=& C_pl_p\exp\left(-\frac{b_{ij}}{l_p}\right)\\
  b_{ij} &=& \frac{1}{2}\sqrt{(\|\vec{d}_{ij}\|+\|\vec{d}_{ij}-\vec{y}_{ij}\|)^{2}-\|\vec{y}_{ij}\|^{2}},
\end{eqnarray}
where $C_p$ and $l_p$ denote the strength and the range of repulsive interaction between pedestrians. $b_{ij}$ denotes the effective distance between pedestrians $i$ and $j$ by considering their relative displacement $\vec{y}_{ij}\equiv (\vec{v}_{j}-\vec{v}_{i})\Delta t$ with the stride time $\Delta t$~\cite{Johansson_2008}. The interpersonal friction $\vec g_{ij}$ becomes effective when the distance $d_{ij}=\|\vec d_{ij}\|$ is smaller than the sum $r_{ij} = r_{i}+r_{j}$ of their radii $r_{i}$ and $r_{j}$:
\begin{eqnarray}\label{eq:friction}
  \vec{g}_{ij} = h(r_{ij}-d_{ij}) \left\{k_n \vec{e}_{ij}+k_t [(\vec{v}_j-\vec v_i)\cdot \vec{t}_{ij}]\vec{t}_{ij}\right\},
\end{eqnarray}
where $k_n$ and $k_t$ are the normal and tangential elastic constants. $\vec{e}_{ij}$ is an unit vector pointing from pedestrian $j$ to $i$ and $\vec{t}_{ij}$ is an unit vector perpendicular to $\vec{e}_{ij}$. The function $h(x)$ yields $x$ if $x > 0$, while it gives $0$ if $x \leq 0$. The repulsive force from boundaries is 
\begin{eqnarray}\label{eq:boundary}
\vec{f}_{iB} = C_b\exp\left(-\frac{d_{iB}}{l_{b}}\right)\vec{e}_{iB},
\end{eqnarray}
where $d_{iB}$ is the perpendicular distance between pedestrian $i$ and wall, and $\vec{e}_{iB}$ is the unit vector pointing from the wall $B$ to the pedestrian $i$. $C_b$ and $l_b$ denote the strength and the range of repulsive interaction from boundaries.

The attractive force toward attractions is modeled similarly to the interaction between pedestrians and the wall, but in terms of both attractive and repulsive interactions. The repulsive effect of attractions is necessary so that pedestrians can keep certain distance from attractions. The strength and range of attractive force are modeled as~\cite{Mogilner_MathBio2003, DOrsogna_PRL2006}
\begin{eqnarray}\label{eq:AR}
\vec{f}_{iA} = \left[C_r\exp\left(\frac{r_i-d_{iA}}{l_r}\right)-C_a\exp\left(\frac{r_i-d_{iA}}{l_a}\right)\right]\vec{e}_{iA}.
\end{eqnarray}
Here $C_a$ and $l_a$ denote the strength and the range of attractive interaction toward attractions, respectively. Similarly, $C_r$ and $l_r$ denote the strength and the range of repulsive interaction from attractions. $d_{iA}$ is the distance between pedestrian $i$ and attraction $A$. $\vec{e}_{iA}$ is the unit vector pointing from attraction $A$ to pedestrian $i$. This study considers short-range strong repulsive and long-range weak attractive interactions, i.e., $C_r > C_a$ and $l_a > l_r$. By these conditions, the attractions attract distant pedestrians but not too close to the attractions.

The interaction between pedestrians and attractions is qualitatively different from others, in that a pedestrian can make decisions between moving in a desired direction and dropping by the attraction, or among multiple attractions. This behavior can be called switching. Such switching behavior can be effectively considered in our setup with plausible values of model parameters, instead of explicitly switching on/off related forces in Eq.~(\ref{eq:EoM}).  

\section{Results and Discussion}
\label{sec:Results}
\subsection{Numerical simulation setup}
\begin{figure}[!t]
\includegraphics[width=.7\columnwidth]{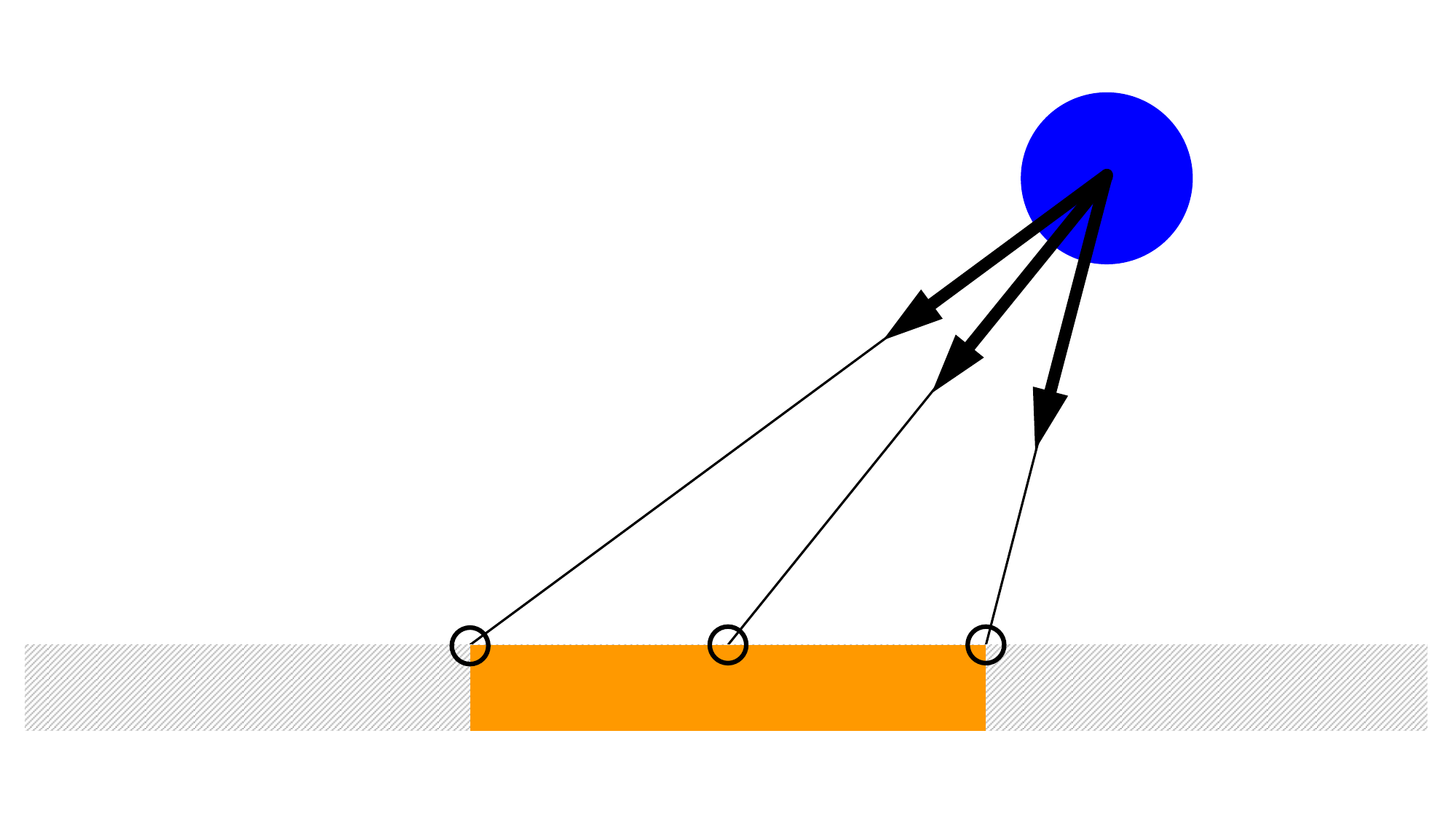}
\caption{(Color online) Each attraction, represented by an orange rectangle, is modeled as three point masses acting on a pedestrian, represented by a blue circle.}
\label{fig:ARpoint}
\end{figure}

In order to investigate the decisive role of attractive interactions toward attractions, numerical simulations are performed. Each pedestrian is modeled by a circle with radius $r_i=0.2$ m. $N$ pedestrians move in a corridor of length 25 m, along the $x$-axis, and width 4 m, along the $y$-axis, with periodic boundary condition in the direction of $x$-axis. They move with desired speed $v_d=1.2$ m/s and with relaxation time $\tau=0.5$ s, and their speed is limited to $v_{\rm max}$ = 2.0 m/s. The desired direction is set to $\vec e_i=(1,0)$ for one half of population and $\vec e_i=(-1,0)$ for the other half. On both the upper and lower walls of corridor, five attractions are evenly spaced, meaning that the distance between attractions is 5 m. To consider the dimension of attractions, each attraction is modeled as three point masses: one point at the center, the other two points at the distance of 0.5 m from the center (See Fig.~\ref{fig:ARpoint}). 

To implement the social force model described in Sec.~\ref{sec:model}, the parameter values in Table~\ref{table-parameter} are selected based on the previous works~\cite{Helbing_PRE1995,Helbing_TrSci2005,Johansson_2008,Moussaid_PNAS2011}. Here the values of $C_r$ and $l_r$ are set to be the same as $C_b$ and $l_b$. In particular, the values of $C_a$ and $l_a$ are chosen such that $l_a>l_r$ and $C_a<C_r$. The effect of attractive interaction toward attractions can be studied by controlling the ratio of $C_a$ to $C_r$, defining the relative attraction strength $C=C_a/C_r$ with $C_r=10.0$. For the corridor area, $A=100$ m$^2$, this study also controls the pedestrian density defined by $\rho=N/A$. At the initial time $t=0$, pedestrians are randomly distributed in the corridor without overlapping. 

\begin{table}[!t]
\caption{Model parameters}
\label{table-parameter}
\begin{tabular}{l*{3}{c}}
\hline\hline 
Model parameter & symbol & value \\
\hline
strength of interpersonal repulsion & $C_p$ & 3.0 \\
range of interpersonal repulsion & $l_p$ & 0.2 \\
normal elastic constants & $k_n$ & 25.0 \\
tangential elastic constants & $k_t$ & 12.5 \\
strength of boundary repulsion & $C_b$ & 10.0 \\
range of boundary repulsion & $l_b$ & 0.2 \\
strength of repulsive interaction from attraction & $C_r$ & 10.0 \\
range of repulsive interaction from attraction & $l_r$ & 0.2 \\
range of attractive interaction toward attraction & $l_a$ & 1.0 \\
pedestrian stride time & $\Delta t$ & 0.5 \\
\hline\hline
\end{tabular}
\end{table}

\subsection{Phase diagram}
\begin{figure}[!t]
\includegraphics[width=\columnwidth]{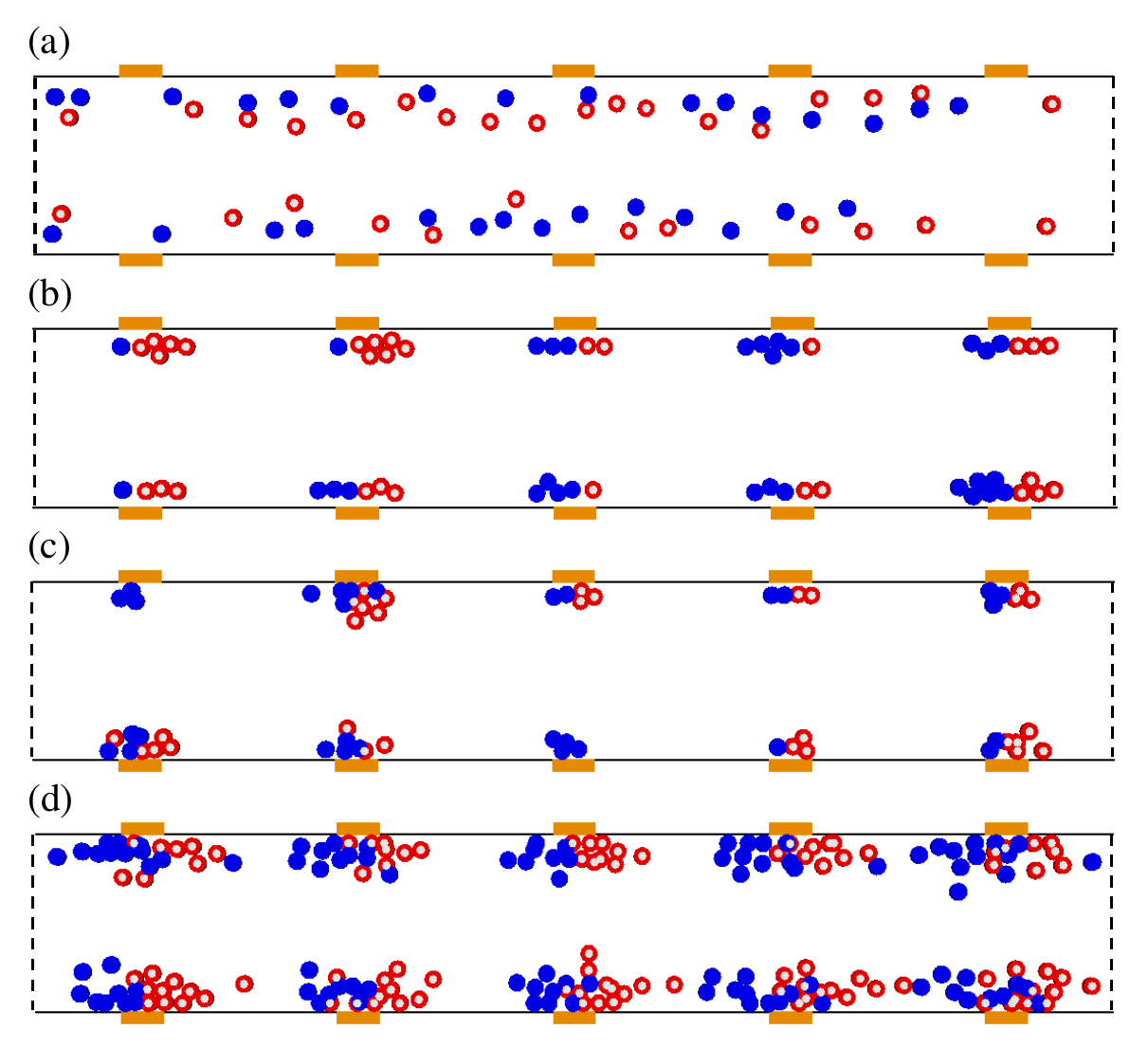}
\caption{(Color online) Snapshots of numerical simulations for various collective patterns, depending on the relative attraction strength $C$ and the pedestrian density $\rho$. The attractions, depicted by rectangles, are located on the walls of the corridor with periodic boundary conditions in the horizontal direction. Filled blue and hollow red circles depict the pedestrians with desired directions to the left and to the right, respectively. The different phases are observed: (a) Free moving phase in the case of $C=0.2$ and $\rho=0.6$, in which pedestrians walk in their desired directions. (b) Agglomerate phase in the case of $C=0.45$ and $\rho=0.6$, in which pedestrians reach a standstill by forming clusters around attractions. (c) Competitive phase in the case of $C=0.7$ and $\rho=0.6$, in which pedestrians rush into attractions and tend to push others. (d) Coexistence subphase in the case of $C=0.55$ and $\rho=2.0$, in which the free moving and the competitive phases coexist.}
\label{fig:snapshots}
\end{figure}

The simulation results show various collective patterns of pedestrian movements depending on relative attraction strength $C$ and pedestrian density $\rho$. When the density is low, three different phases are observed depending on the relative attraction strength: (i) When the attractive interaction $C$ is small, pedestrians walk in their desired directions, which enables to define the free moving phase, see Fig.~\ref{fig:snapshots}(a). (ii) For the intermediate range of $C$, one can observe an agglomerate phase, where pedestrians reach a standstill by forming clusters around attractions, although they intended to walk with their desired velocity [See Fig.~\ref{fig:snapshots}(b)]. (iii) The competitive phase appears when $C$ is large. Since increasing the relative attraction strength leads to extreme desire for attractions, pedestrians rush into attractions and tend to push others as shown in Fig.~\ref{fig:snapshots}(c). On the other hand, if the density is high, the agglomerate phase is not observed any more, while the free moving and competitive phases coexist only for the intermediate range of $C$ as shown in Fig.~\ref{fig:snapshots}(d).

In order to quantitatively distinguish different pedestrian flow patterns, this study employs the efficiency of motion $E$ and the normalized kinetic energy $K$~\cite{Helbing_PRL2000} as follows:
\begin{eqnarray}\label{eq:E}
E = \left\langle \frac{1}{N} \sum_{i=1}^{N} \frac{\vec{v}_{i}\cdot\vec{e}_{i}}{v_d} \right\rangle
\end{eqnarray}
and
\begin{eqnarray}\label{eq:K}
K = \left\langle \frac{1}{N} \sum_{i=1}^{N} \frac{\|\vec{v}_{i}\|^2}{v_d^2} \right\rangle.
\end{eqnarray}
Here $\langle\cdot\rangle$ represents an average over 60 independent simulation runs after reaching the stationary state. The efficiency reflects the contribution of the driving force in the pedestrian motion. If all pedestrians walk with their desired velocity, the efficiency becomes $1$. On the other hand, the zero efficiency can be obtained if pedestrians do not move in their desired directions and form clusters at attractions. The normalized kinetic energy has the value of $0$ if all pedestrians do not move, otherwise it has a positive value.

\begin{figure}
\includegraphics[width=\columnwidth]{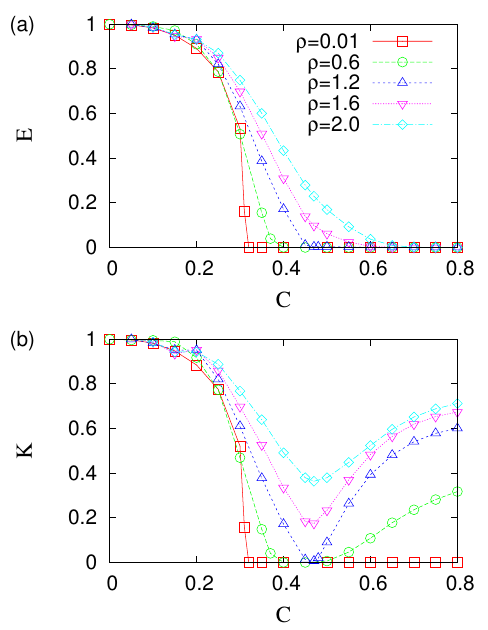}
\caption{(Color online) Numerical results of the efficiency of motion $E(C,\rho)$ (a) and the normalized kinetic energy $K(C,\rho)$ (b). Different symbols represent the different values of $\rho$. For each given $\rho$, $E(C)$ monotonically decreases according to $C$, while $K(C)$ decreases and then increases except for small $\rho$. Different phases and transitions among them can be characterized in terms of the behaviors of $E$ and $K$.}
\label{fig:graph}
\end{figure}

Figure~\ref{fig:graph} shows how the efficiency $E(C,\rho)$ and the normalized kinetic energy $K(C,\rho)$ depend on the pedestrian density $\rho$ and the relative attraction strength $C$. For a given $\rho$, $E$ decreases according to $C$, meaning that pedestrians are more distracted from their desired velocity due to the larger strength of attractions. Interestingly, $E$ becomes $0$ at a finite value of $C=C_0(\rho)$, indicating the transition from the free moving phase to the agglomerate phase for low $\rho$ or to the competitive phase for high $\rho$. In case with $N=1$, . $\rho=0.01$, the pedestrian movements end up with either moving or stopping the movement, depending on $C$. That is, the transition for $N=1$ is discontinuous, depicted as an abrupt change in the curve of $\rho=0.01$ in Fig.~\ref{fig:graph}(a). The same behavior is observed up to $\rho=0.1$, i.e., one pedestrian per attraction on average. 

For the region of $\rho>0.1$, the decreasing behavior of $E$ becomes continuous mainly due to the collective effect of interpersonal repulsion. In addition, the larger $E$ is observed with higher $\rho$, which can be explained by the interpersonal repulsion effect. The pedestrians moving far from the walls become less distracted by the attractions because of the repulsion by pedestrians close to the attractions. This interpersonal repulsion also explains why the transition point of $C_0(\rho)$ is larger for the higher $\rho$. The stronger attraction is needed to attract distant pedestrians.

Although the efficiency enables to identify the boundary of the free moving phase, the properties of agglomerate and competitive phases can be better understood in terms of the normalized kinetic energy. The decreasing behavior of $K$ is overall similar to that of $E$ up to $C\approx 0.45$. It also turns out that $K$ begins to increase for larger values of $C$ except for $\rho<0.1$. In the case of $\rho<0.1$, since pedestrians do not occupy the same attraction on average, the further increasing strength of attractions do not change pedestrian movements, keeping $K=0$. The increasing behavior of $K$ is of two kinds, depending on the range of $\rho$. First, the case with low values of $\rho$ is considered. As $C$ increases, $K$ turns out to be $0$ at the same critical point of $C=C_0$, and then gradually increases from $0$ at $C=C_+(\rho)$ with $C_+\geq C_0$ (see Fig.~\ref{fig:graph}). The boundaries among different phases can be identified in terms of $C_0$ and $C_+$. The free moving phase for $C<C_0$ is characterized by 
\begin{equation}
  E>0\ \textrm{and}\ K>0,
\end{equation}
and the agglomerate phase for $C_0<C<C_+$ by 
\begin{equation}
 E=K=0. 
\end{equation}
When $C>C_+$, 
\begin{equation}
  E=0\ \textrm{and}\ K>0
\end{equation}
characterizes the competitive phase, in which pedestrians rush into attractions and jostle each other in order to get closer to attractions. This also explains the observation of $\partial K/\partial C>0$. Even though the stronger attractions intensify jostling behavior of pedestrians, pedestrians do not jump to the neighboring attractions. Moreover, due to the competition, they stay near attractions and frequently move in the opposite direction of the desired velocity, leading to $E=0$. It is found that $C_0(\rho)$ increases and $C_+(\rho)$ decreases according to the increasing $\rho$. They finally coincide at the crossover density $\rho_\times\approx 1.22$, in which the agglomerate phase vanishes. The value of $\rho_\times$ indicates the maximum number of pedestrians that can stay near attractions without jostling and can be interpreted as a total capacity of attractions. In this simulation setup, each attraction can accommodate about up to 12 pedestrians on average. 

\begin{figure}
\includegraphics[width=\columnwidth]{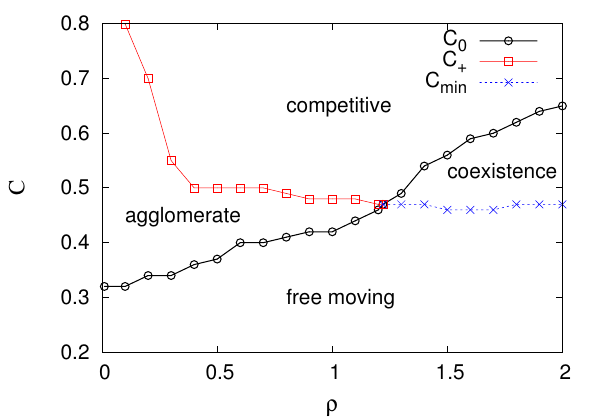}
\caption{(Color online) Phase diagram summarizing the numerical results. The parameter space of the pedestrian density $\rho$ and the relative attraction strength $C$ is divided into four regions by means of $C_0$ ($\circ$), $C_+$ ($\Box$), and $C_{\rm min}$ ($\times$).}
\label{fig:diagram}
\end{figure}

Secondly, for the high values of $\rho>\rho_\times$, $K$ decreases and then increases according to $C$ but without becoming zero. The striking difference from the case with low $\rho$ is the existence of parameter region characterized by both 
\begin{equation}
  E>0\ \textrm{and}\ \frac{\partial K}{\partial C}>0.
\end{equation}
This implies that some pedestrians rush into attractions as in the competitive phase ($\partial K/\partial C>0$), while other pedestrians move in their desired directions as in the free moving phase ($E>0$). The moving pedestrians are also attracted to attractions but cannot stay around them because of interpersonal repulsion effect by other pedestrians closer to attractions. Thus, this parameter region can be called coexistence subphase. The coexistence subphase belongs to the free moving phase in the sense that the coexistence subphase is also characterized by $E>0$ and $K>0$. However, one can identify the lower boundary of coexistence subphase by determining $C_{\rm min}$ that minimizes $K$. At $C_{\rm min}$, the number of pedestrians accommodated by attractions is also maximized. $C_{\rm min}$ seems to be independent of $\rho$. It is because the excessive pedestrians proportional to $\rho-\rho_\times$ cannot be accommodated by attractions, hence move between attractions, which elevates the curve of $K$ at least around $C_{\rm min}$. The upper boundary to the competitive phase is determined by the critical point $C_0$. It is found that $C_0(\rho)$ is an increasing function of $\rho$ because stronger attractions are needed to entice more pedestrians. Different phases characterizing different collective patterns of pedestrians and transitions among them are summarized in the phase diagram of Fig.~\ref{fig:diagram}.

\subsection{Spatiotemporal patterns}
\begin{figure}
	\includegraphics[width=\columnwidth]{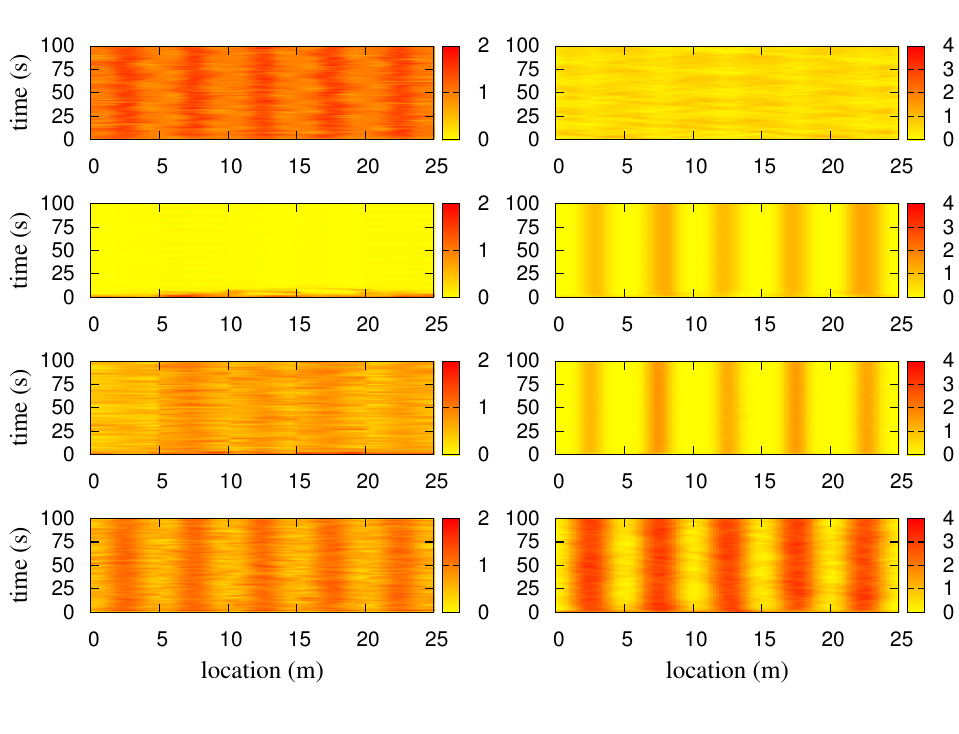}
	\caption{(Color online) Local speed maps (left) and local density maps (right) for various collective patterns of pedestrian movements: free moving, agglomerate, and competitive phases, and coexistence subphase (from top to bottom). Yellow (light) and red (dark) colors indicate lower and higher values, respectively.}
	\label{fig:local}
\end{figure}

In addition to the global quantities characterizing the collective patterns, such as efficiency of motion and normalized kinetic energy, one can better describe the patterns by means of local quantities. As in~\cite{Helbing_PRE2007,Moussaid_PNAS2011}, the local speed and the local density are adopted to describe the spatiotemporal patterns of pedestrian movements. The local speed at a location $\vec z$ and time $t$ is measured as 
\begin{eqnarray}\label{eq:LocalSpeed}
V(\vec{z}, t) = \frac{\sum_{i}^{ }{\|\vec{v}_{i}\|f(d_{iz})}}{\sum_{i}^{ }{f(d_{iz})}},
\end{eqnarray}
where $d_{iz}$ is the distance between location $\vec{z}$ and pedestrian $i$'s position. $f(d)$ is a Gaussian distance-dependent weight function
\begin{eqnarray}\label{eq:Gauss}
f(d) = \frac{1}{\pi R^2}\exp\left(-\frac{d^2}{R^2}\right)
\end{eqnarray}
with a measurement parameter $R = 0.7$. The local density is similarly defined as
\begin{eqnarray}\label{eq:LocalDensity}
\rho(\vec{z}, t) = \sum_{i}^{ }{f(d_{iz})}.
\end{eqnarray}
Figure~\ref{fig:local} shows for different phases the local speed maps $V(x,t)$ and the local density maps $\rho(x,t)$ that have been averaged along the $y$-axis. For the free moving phase, pedestrians speed up when they approach the attractions, and then slow down when moving away from them. The fast moving pedestrians around attractions lead to the low density, while slowly moving pedestrians between attractions is related to the high density. In the agglomerate phase, pedestrians form clusters around attractions, resulting in the high local density around attractions and zero local speed in the entire corridor. In the competitive phase, the local density shows tighter clustering formation around attractions than that of the agglomerate phase. One can observe fluctuations in the local speed in that frenzied pedestrians rush into attractions. The coexistence subphase shows a similar pattern to the free moving phase in the local speed map and to the competitive phase in the local density map.

\section{Conclusion}
\label{sec:summary}
This study has numerically investigated collective effects of attractive interactions between pedestrians and attractions by extending the social force model. A phase diagram with various collective patterns of pedestrian movements is presented. The phases are identified by means of the efficiency of motion $E$ and the normalized kinetic energy $K$ as functions of the pedestrian density $\rho$ and the relative attraction strength $C$. For low density, as $C$ increases, the transition occurs from the free moving phase ($E,K>0$) to the agglomerate phase ($E=K=0$) and finally to the competitive phase ($E=0$ and $K>0$). For high density, the agglomerate phase disappears and a coexistence subphase emerges where the properties of free moving and competitive phases are observed simultaneously for the intermediate range of $C$. The crossover density separating low and high densities can be understood as the total capacity of attractions. It is noted that these results are robust with respect to the choice of relevant parameter values for attractive interactions, i.e., $l_a$ and $C_r$. 


The findings from this study can provide insight into customer behavior in stores. The driving force and the attractive force can be interpreted as the initial plan and the influence of unexpected attractions, respectively. Accordingly, the appearance of the agglomerate phase might be analogous to consumer impulse purchase without an initial plan~\cite{Rook_1987, Rook_1995, Baumeister_2002}. The competitive phase can be relevant to understanding the pedestrian incidents involving extreme desire for limited resources. This study also suggests that safe and efficient use of pedestrian facilities can be achieved by moderating the density and the strength of attractive interaction. 

A very simple scenario has been focused in order to study the fundamental role of attractive interactions. Assuming superposition of various interaction terms in the social force model could yield switching behavior in effect by choosing plausible parameter values. However, switching behavior can be explicitly formulated for more realistic models, and should be verified against experimental data. Indeed, it is widely known that people have limited amount of attention, so they tend to focus on few stimuli rather than consider every stimulus around them~\cite{Wickens_1999, Goldstein_2007}. In addition, one can take into account heterogeneous properties of pedestrians and attractions, such as different strength of attractive interactions between pedestrians and attractions.
\\
\\
\begin{acknowledgments}
This work is funded by Aalto University 4D-Space MIDE project (JK) and by Aalto University postdoctoral program (HJ).
\end{acknowledgments}



\end{document}